\documentclass[aps,pre,preprint]{revtex4-1}

\usepackage{setspace}
\usepackage{amstext}
\usepackage{amsmath}%
\usepackage{amsfonts}%
\usepackage{amssymb}%
\usepackage{amsbsy}
\usepackage{amsgen}

\usepackage{color}

\usepackage{graphicx}

\newlength{\fw}
\begin{document}
\title{Directional locking in deterministic lateral displacement microfluidic separation systems}

\author{Sumedh R. Risbud}
\affiliation{Chemical and Biomolecular Engineering, Johns Hopkins University}
\author{German Drazer}
\affiliation{Mechanical and Aerospace Engineering, Rutgers, the State University of New Jersey}

\begin{abstract}
We analyze the trajectory of suspended spherical particles moving through a square array of obstacles, in the deterministic limit and at zero Reynolds number. 
We show that, in the dilute approximation of widely separated obstacles, the average motion of the particles is equivalent to the trajectory followed by a point 
particle moving through an array of obstacles with an effective radius. The effective radius accounts for the hydrodynamic as well as short-range repulsive 
non-hydrodynamic interactions between the suspended particles and the obstacles and is equal to the critical offset at which particle trajectories become irreversible. 
Using this equivalent system we demonstrate the presence of directional locking in the trajectory of the particles and derive an inequality that accurately describes the
{\em Devil's staircase} type of structure observed in the migration angle as a function of the forcing direction. Finally, we use these results to determine the
optimum resolution in the fractionation of binary mixtures using deterministic lateral displacement separation microfluidic systems.
\end{abstract}

\maketitle

\section{Introduction}\label{intro}
One of the essential unit operations in micro-total-analysis-systems ($\mu$TAS) is the separation of 
species for downstream analysis. Early microfluidic separation strategies involved miniaturization of 
different macroscopic separation methods, e.g., size exclusion \citep{chirica2006} and hydrodynamic 
chromatography \citep{blom2003}. However, current micro-fabrication techniques enable design and 
fabrication of precisely controlled micro-structures to act as separation media, in contrast with the 
random micro-structure common in conventional separation media. For example, `entropic trapping' 
a channel with alternating thick and thin regions was used to separate DNA molecules by size based 
on the time time they spend in the entropic traps (thick regions) \citep{han2000}. 
In `pinched flow fractionation', species entering a constriction and exiting into a sudden expansion experience a 
lateral displacement from their trajectories that is a function of their size \citep{yamada2004}. 
`Deterministic lateral displacement' (DLD) employs a periodic array of solid obstacles, through which species 
of different sizes migrate in different spatial directions in the presence of the same driving force \citep{huang2004} . 
This effect can also be achieved with a periodic array of optical traps (soft potentials instead of 
solid obstacles, \citep{macdonald2003}). Although DLD systems have been studied extensively 
\citep{huang2004,inglis2008,korda2002,lacasta2005,gopinathan2004}, the understanding of the underlying 
mechanism is presented only heuristically, and lacks a theoretical framework for their analysis.

We have performed numerous detailed computational and experimental studies of DLD-like systems 
\citep{balvin2009,frechette2009,herrmann2009,koplik2010,bowman2012,devendra2012} -- where the 
experiments include microfluidic as well as macroscopic platforms at low Reynolds number -- 
and have established that {\em directional locking} dictates the particle trajectories in such 
systems.  In this work, we focus on the mechanism underlying separations in the DLD systems. 
Specifically, we present a theoretical analysis of DLD 
systems, involving the motion of a particle of arbitrary radius in a square array of obstacles of 
circular cross-section. We assume a `dilute limit' for the obstacles, such that the inter-obstacle 
spacing is sufficiently large and a particle interacts with a single obstacle at a time. The field 
driving the particle (either a constant force, or a flow field) is assumed to be at an arbitrary angle 
(henceforth, forcing angle) with respect to the principal lattice directions of the square array. We 
assume negligible particle as well as fluid inertia, and infinite P\'{e}clet number (non-Brownian 
particles, deterministic trajectories). We show that, under the dilute approximation,
the particle-obstacle interaction can be replaced by a point particle moving in straight lines past 
an obstacle with an effective radius equal to the critical offset.
The critical offset is the offset at which particle-obstacles collisions become irreversible and can be
interpreted as a length-scale that characterizes the effect of short-range repulsive non-hydrodynamic 
interactions between the particles and the obstacles. (Previous work presents a detailed discussion of
the critical offset, both computationally \citep{frechette2009} as well as theoretically \citep{risbud2013}) . 
Using this equivalent representation of the system under the dilute 
approximation, the problem of calculating the particle trajectories reduces to simple geometric 
manipulations. We derive a periodicity criterion for particle trajectories in terms of the design 
parameters of the system, namely, the critical offset, the forcing angle, and the inter-obstacle 
spacing in the square array. The periodicity criterion yields the experimentally and computationally 
observed directional locking behavior. Further, we show that the same framework can be used to uncover 
size-based spatial band-pass filtering of particles through the array, such that particles of a 
certain intermediate size migrate with an average migration angle larger than that corresponding to 
smaller as well as larger particles.

The article is organized as follows: in \S\ref{sec:sysDefDLD} we introduce the system under consideration, 
the system variables, and the dilute approximation. We also explain the model for short-range repulsive 
non-hydrodynamic interactions leading to the definition of the critical parameter $b_c$, and establish an 
abstract model for the particle-obstacle pair. In \S\ref{sec:dirLock}, we use the abstract model to derive 
a periodicity condition for particle trajectories. We apply the periodicity condition to derive expressions 
for the simplest locking directions in \S\ref{subsubsec:firstLockDirP} and \S\ref{subsubsec:lastLockDirQ}. 
In \S\ref{sec:designRules}, we use the periodicity condition corresponding to particles exhibiting the 
simplest directional locking behavior, and comment on the resolution of separation between such particles. 
%
\begin{figure*}
\begin{center}
\includegraphics[width=1.25\fw]{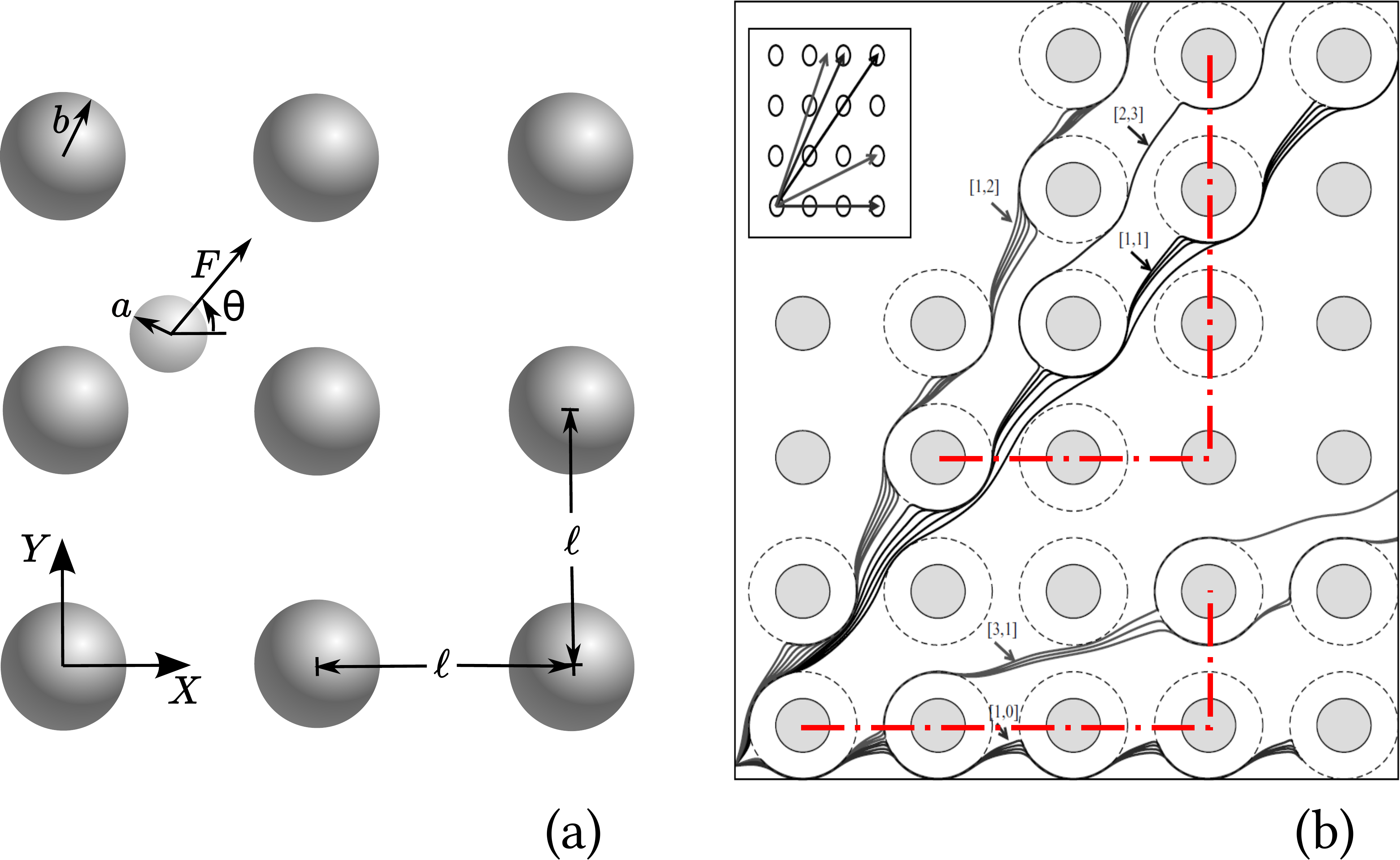}
\end{center}
\caption{(a) A spherical particle of radius $a$ negotiating a portion of a square array of obstacles of circular 
cross-section with radius $b$ [adapted from \citep{frechette2009}]:
the length of a unit-cell is $\ell$, the driving field $\boldsymbol{\mathit{F}}$, oriented at an angle $\theta$ as 
shown, drives the particle through the array. The principal lattice-directions are indicated with Cartesian axes 
$X$ and $Y$. (b) A few example particle trajectories exhibiting directional locking [adapted from 
\citep{frechette2009}]: results of Stokesian dynamics simulations with $a=b$, $\ell=5a$ and the range of 
non-hydrodynamics interactions $\epsilon=10^{-3}$ (see \S\ref{sec:sysDefDLD} for a discussion on non-hydrodynamic 
interactions). Counter-clockwise, from $X$-axis to $Y$-axis, the trajectories can be seen to be locked in 
directions $[1,0]$, $[3,1]$, $[1,1]$, $[2,3]$ and $[1,2]$ (the inset shows the migration directions). The 
dot-dashed lines are to guide the eye and highlight $[3,1]$ and $[2,3]$ locking directions.}
\label{fig:systemGeometryDLD}
\end{figure*}

\section{System description, assumptions and abstractions}\label{sec:sysDefDLD}
Figure \ref{fig:systemGeometryDLD} depicts the system under investigation. We consider a suspended spherical 
particle of radius $a$ negotiating a square array of obstacles with circular cross section of radius $b$, 
under the action 
of a driving field $\boldsymbol{\mathit{F}}$ (either a constant force, or a uniform flow away from the lattice). 
The field is oriented at an angle $\theta$ with respect to one of the principal axes of the array (say, the 
$X$-axis as shown in the figure). The lattice spacing is $\ell$. The domain of the `forcing angle' is restricted 
to $\theta\in[0,\frac{\pi}{4}]$, since the system possesses a reflection symmetry in the $X=Y$ line.

\begin{figure*}
\begin{center}
 \includegraphics[width=1.25\fw]{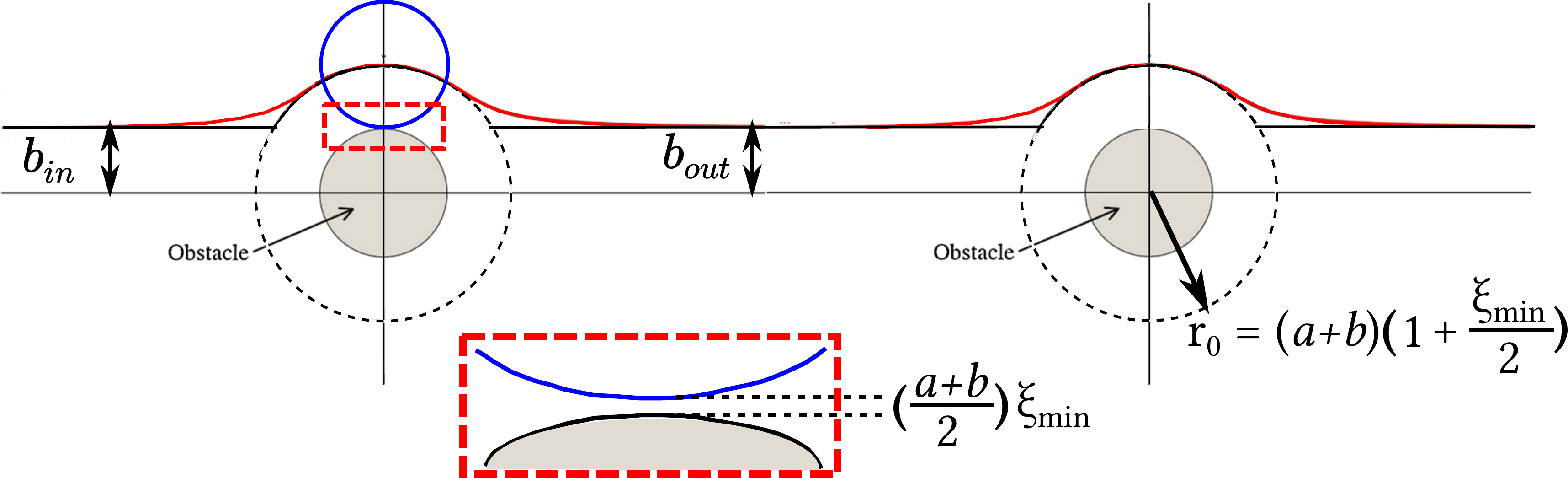}
\end{center}
 \caption{A schematic depicting the variables of the problem, $b_{in},~b_{out}$ and $\xi_{min}$. The dashed 
circular region (excluded volume) has radius $r_0 = (a+b)(1+\frac{\xi_{min}}{2})$. The schematic qualitatively 
shows the `dilute approximation', wherein two consecutive obstacles are sufficiently separated that 
considering only the interaction between the particle and its closest obstacle is a good approximation. Note that 
the particle trajectory is seen to attain the asymptotic value of $b_{out}$ in between consecutive collisions.}
 \label{fig:compareDiluteToExact}
\end{figure*}

We work in the `Stokes regime', i.e., we neglect fluid inertia (vanishingly small Reynolds number) and particle 
inertia (vanishingly small Stokes number). We consider the deterministic limit (infinitely large P\'{e}clet number, 
non-Brownian limit). Further, in DLD micro-devices, the enclosing walls perpendicular to the $Z$-axis (i.e., 
walls parallel to the plane of the paper) screen the hydrodynamic interactions between the particle and distant 
obstacles. Therefore, we assume that the lattice spacing $\ell$ is sufficiently larger than the $Z$-spacing between 
the walls, such that, to a good approximation we can consider the interaction between the particle and only 
the closest obstacle (figure \ref{fig:compareDiluteToExact}, dilute approximation). 
Figure \ref{fig:compareDiluteToExact} depicts the variables of the problem; the incoming and outgoing offsets 
are denoted by $b_{in}$ and $b_{out}$, respectively. The dimensionless minimum surface-to-surface  
separation attained by the particle from the obstacle is denoted by $\xi_{min}$ in the figure. The functional 
relationship between $b_{in}$ and $\xi_{min}$ explicitly incorporates the hydrodynamic mobility of the particle 
around the fixed obstacle, thereby taking into consideration the hydrodynamic interactions \citep{risbud2013}.

\begin{figure}
\begin{center}
 \includegraphics[width=0.85\fw]{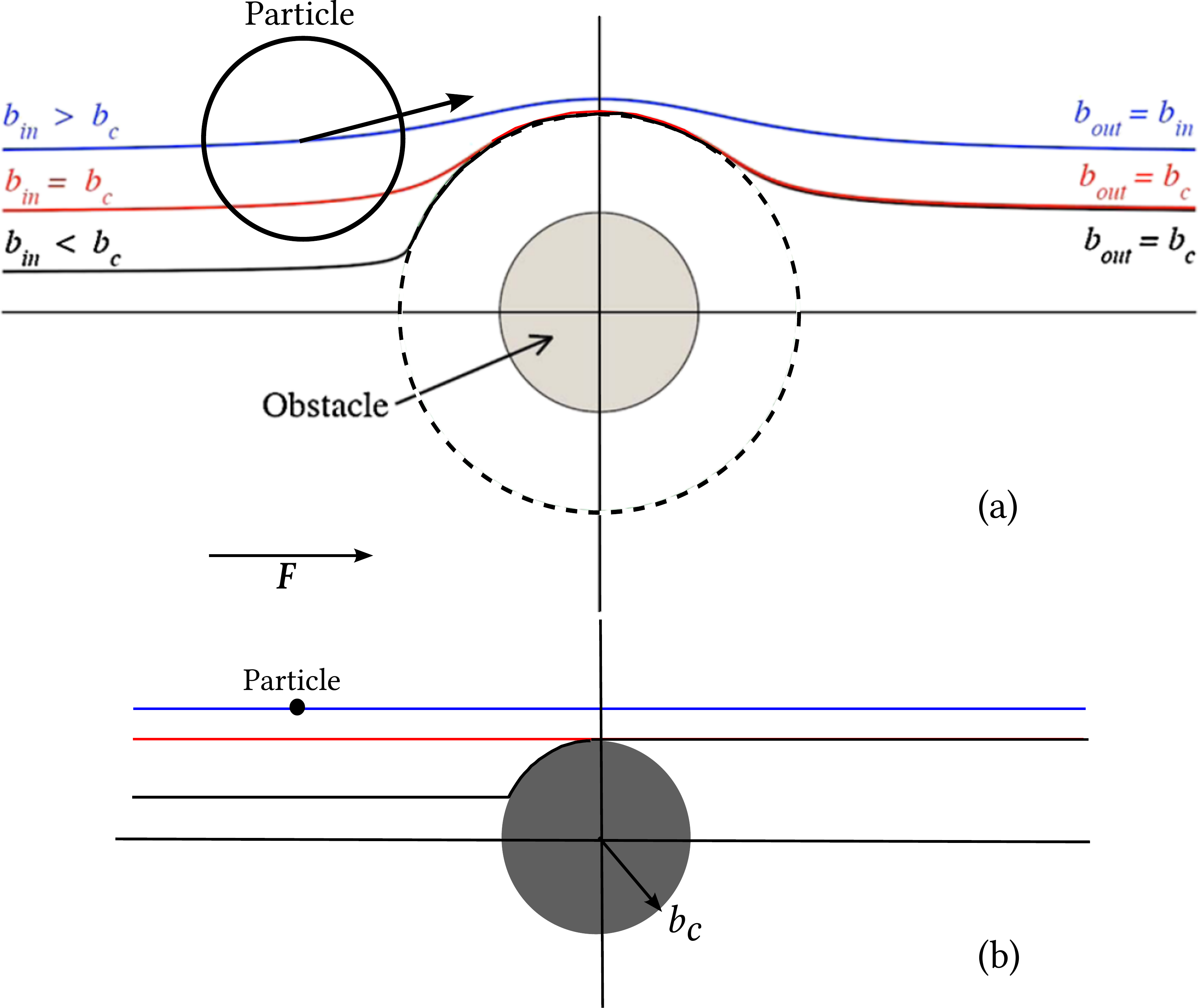}
\end{center}
 \caption{(a)[Adapted from \citep{balvin2009}] Three kinds of particle trajectories in the presence of short-range 
repulsive non-hydrodynamic interactions. (b) Depiction of the equivalent system in which a point-particle traverses 
past an obstacle of radius $b_c$ (effective radius). The outgoing part of the trajectories 
with $b_{in}<b_c$ is tangent to the obstacle in the equivalent system.}
 \label{fig:threeTrajAndAbstraction}
\end{figure}

Apart from the hydrodynamic interactions between the particle and the obstacle that arise from their finite size, 
we also take into account the effect of short-range repulsive non-hydrodynamic interactions such as solid-solid 
contact due to surface roughness, electrostatic repulsion, steric repulsion, etc. A simple and effective model 
for these non-hydrodynamic interactions is to treat them as leading to a hard-wall potential with a given 
dimensionless range $\epsilon$, such that it creates a hard shell around the obstacle and the particle surface 
cannot approach the obstacle surface closer than $\epsilon$ 
\citep{davis1992b,dacunha1996,rampall1997,brady1997,wilson2000,bergenholtz2002,drazer2002,davis2003,drazer2004,ingber2008,frechette2009,blanc2011}. 
We have shown elsewhere that the presence of such non-hydrodynamic repulsion leads to the occurrence of a 
critical offset $b_c$ \citep{risbud2013, risbud2013a}. This can be further elaborated with the aid of figure 
\ref{fig:threeTrajAndAbstraction}(a). As shown, the particle trajectories can be categorized as follows (from 
top to bottom): (a) the trajectories for which $\xi_{min}>\epsilon$. In this case, the particle motion is unaffected 
by the presence of the non-hydrodynamic interactions, (b) the trajectory that corresponds to $\xi_{min}=\epsilon$. 
In this case, the particle `grazes' the obstacle and defines the {\em critical trajectory}, and (c) the trajectories 
that would correspond to $\xi_{min}<\epsilon$ in the absence of non-hydrodynamic interactions. However, in this case, 
the particles are forced to circumnavigate the obstacle by maintaining a constant separation equal to $\epsilon$ on the 
approaching side due to the hard-core potential. The last group of trajectories collapse onto the critical trajectory 
downstream of the obstacle, breaking their fore-aft symmetry. Thus, the critical trajectory (of type (b) described 
above) defines the critical offset as $b_{in} = b_c$, such that the corresponding minimum separation is 
the range of the non-hydrodynamic interactions (i.e., $\xi_{min,c}=\epsilon$). Therefore, in the 
presence of short-range repulsive non-hydrodynamic interactions, the relationship between $b_{in}$ and $\xi_{min}$ 
is equivalent to the relationship between the critical offset $b_c$ and the range of the interactions $\epsilon$ 
\citep{balvin2009,herrmann2009,frechette2009,luo2011,risbud2013,risbud2013a}.

Using the hard-wall model for the non-hydrodynamic interactions combined with the dilute assumption, we can thus 
replace the physical particle-obstacle system with an equivalent abstract system shown in figure 
\ref{fig:threeTrajAndAbstraction}(b). The obstacle radius $b$ can be replaced by $b_c$ and the particle can be reduced 
to a point particle. As shown, since the particle trajectories with incoming impact parameter $b_{in}>b_c$ remain 
fore-aft symmetric, one can replace them with straight lines uninfluenced by the obstacle. The trajectories with 
$b_{in}<b_c$ (that would intersect the new, abstract obstacle), get laterally displaced by $(b_c-b_{in})$, and 
continue as tangents to the obstacle parallel to the forcing direction. It is interesting to note that both 
the hydrodynamic as well as non-hydrodynamic interactions are incorporated in the single parameter $b_c$.

\section{Mathematical description of directional locking}\label{sec:dirLock}
The defining feature of deterministic lateral displacement is directional locking of particle trajectories (see 
figure \ref{fig:systemGeometryDLD}(b)). In a square array of obstacles (e.g., DLD devices), the particle follows 
a periodic trajectory with a periodicity of (say) $p$ lattice units in $X$-direction and $q$ lattice units in $Y$-direction 
for a range of values of $\theta$, and some integers $p$ \& $q$. In such a case, the trajectory is said to be 
locked in the $[p,~q]$ direction for that range of values of $\theta$. The migration angle $\alpha$ is defined by, 
\[\tan\alpha = \frac{q}{p}.\]

\begin{figure*}
\begin{center}
 \includegraphics[width=1.35\fw]{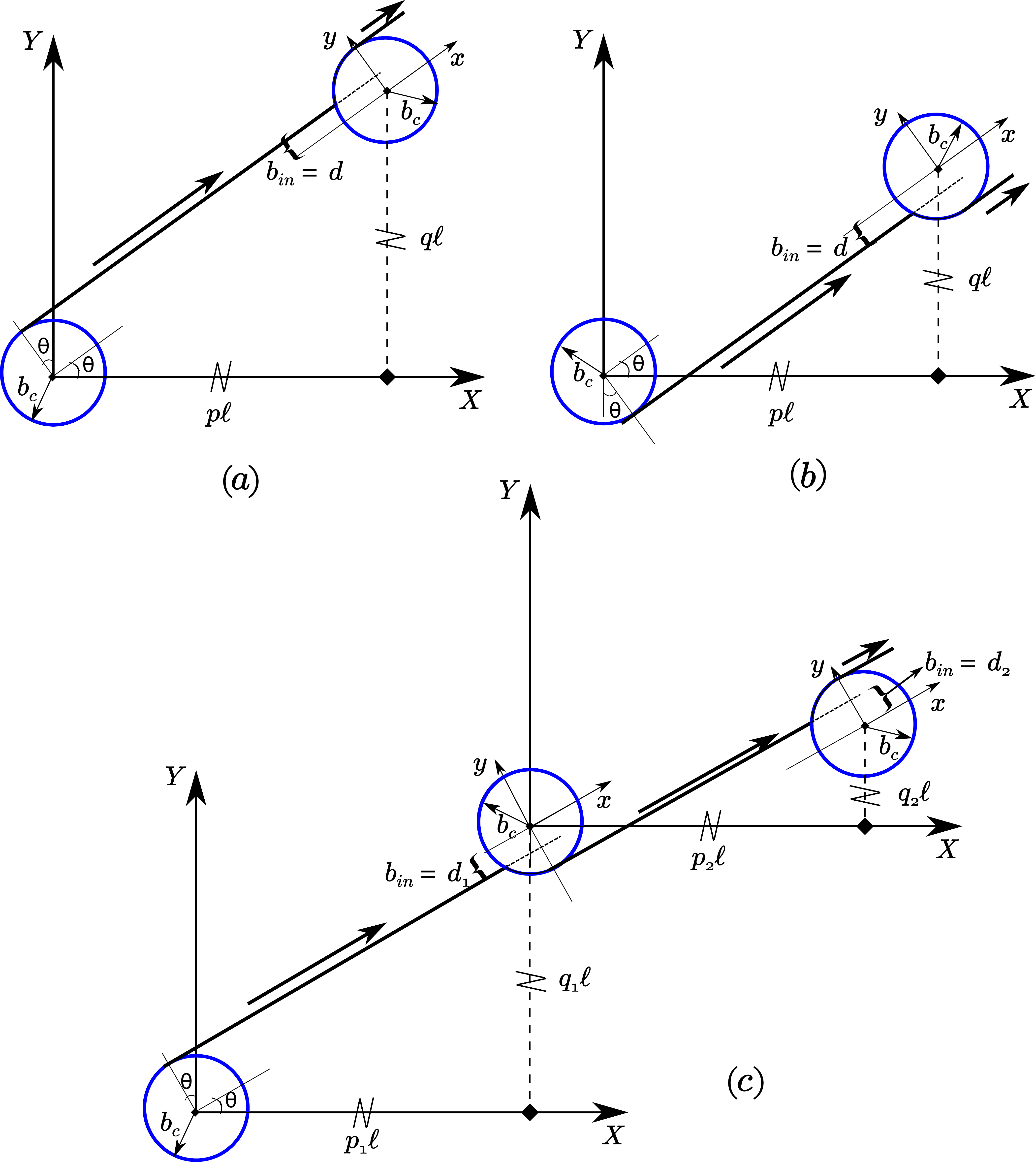}
\end{center}
 \caption{Schematic depicting three possibilities leading to periodic trajectories (see text). In (a) and (b), 
the trajectories repeat after $p$ obstacles along $X$-axis and $q$ obstacles along $Y$-axis. In (c), the period 
along $X$-axis is ($p_1+p_2$) and that along $Y$-axis is $(q_1+q_2)$.}
 \label{fig:deriveSchematic}
\end{figure*}

Equipped with the abstraction of the particle-obstacle pair described in the previous section, we now 
consider a square array of such obstacles with radius $b_c$, separated by the lattice spacing $\ell$. Figure 
\ref{fig:deriveSchematic} shows a schematic of the equivalent system with straight-line trajectories between 
two successive particle-obstacle collisions that occur $p$ lattice units apart in $X$-direction and $q$ lattice units 
apart in $Y$-direction, thereby representing a $[p,q]$-periodic trajectory. The figure shows two coordinate systems, 
the $XY$-system with its axes parallel to the principal axes of the lattice as well as the $x y$-system with 
$x$-axis parallel to the direction of the driving field $\boldsymbol{\mathit{F}}$. Since we have a point-particle 
traversing in a straight line parallel to the direction of the driving field, it is evident that a particle-obstacle 
interaction (a `collision') is possible only if the particle trajectory intersects the obstacle, i.e., only if the 
distance $d$ to the obstacle center from the trajectory is less than the obstacle radius. Note that, as shown in figure 
\ref{fig:deriveSchematic}, $d$ is the same as the initial offset $b_{in}$ for the corresponding obstacle. It is evident 
that there are only two kinds of collisions with respect to the sign of the $y$-coordinate of the point of collision, 
{\em top} ($y>0$) and {\em bottom} ($y<0$) ones. Therefore, a given periodic trajectory, can exhibit periodicity in 
exactly three distinct modes: (a) all successive collisions satisfy $y > 0$ ({\em top-top collisions}, figure 
\ref{fig:deriveSchematic}(a)), (b) all successive collisions satisfy $y < 0$ ({\em bottom-bottom collisions}, figure 
\ref{fig:deriveSchematic}(b)), or (c) collisions alternately satisfy $y > 0$ and $y < 0$ ({\em top-bottom-top collisions} 
or equivalently, {\em bottom-top-bottom collisions} figure \ref{fig:deriveSchematic}(c))

As shown in figure \ref{fig:deriveSchematic}, we choose an arbitrary obstacle, which has undergone a collision, 
as the origin of the $XY$-system. In the case of top-top and bottom-bottom collisions (figure 
\ref{fig:deriveSchematic}(a) and (b)), we assume that the period 
is $p$ in $X$-direction, and $q$ in $Y$-direction, for some integers $p~\&~q$. Hence the coordinates of the center 
of the next obstacle are ($p\ell,~q\ell$) in figures \ref{fig:deriveSchematic}(a) and (b). In the case of periodic 
trajectories arising from top-bottom-top (equivalently, bottom-top-bottom) collisions (figure 
\ref{fig:deriveSchematic}(c)), we assume that $p_1$ and $p_2$ are the alternate periods in $X$-direction, while 
$q_1$ and $q_2$ are the periods in $Y$-direction, again, for some integers $p_1,~p_2,~q_1~\&~q_2$.

\subsection{The periodicity-condition}\label{subsec:derivePeriodicity}
For a top-top collision, the equation of the trajectory in the $XY$-system is, \[Y = X\tan\theta + b_c\sec\theta.\] 
Since the center of the next obstacle, ($p\ell,~q\ell$), lies in the lower half-plane of the trajectory, it 
satisfies $q\ell < p\ell\tan\theta + b_c\sec\theta$. Therefore, the normal distance between the obstacle center 
and the trajectory in the case of top-top collisions is,
\begin{equation}
d_{TT} = \frac{p\ell\tan\theta-q\ell+b_c\sec\theta}{\sqrt{1+\tan^2\theta}} = p\ell\sin\theta - q\ell\cos\theta + b_c
\label{eqn:dTopTop}
\end{equation}
For a top-top collision, in the $x y$-system centered on the second obstacle, the initial offset 
must satisfy $0\le b_{in}=d_{TT}<b_c$. Therefore, (\ref{eqn:dTopTop}) yields,
$ 0< q\ell\cos\theta - p\ell\sin\theta \le b_c$. This inequality can be rephrased as,
\begin{equation}
 0 < \sin\left(\alpha-\theta\right) \le \frac{b_c}{s\ell},
\label{eqn:dTopTopCondn}
\end{equation}
where, $s\left([p,q]\right)=\sqrt{p^2+q^2}$.

A similar procedure for bottom-bottom collisions dictates that the trajectory is described by \[Y = X\tan\theta - 
b_c\sec\theta.\] The obstacle center ($p\ell,~q\ell$), lies in the upper half-plane of the trajectory satisfying 
$q\ell > p\ell\tan\theta - b_c\sec\theta$. Therefore,
\begin{equation}
d_{BB}=\frac{q\ell-p\ell\tan\theta+b_c\sec\theta}{\sqrt{1+\tan^2\theta}}=q\ell\cos\theta - p\ell\sin\theta + b_c
\label{eqn:dBotBot}
\end{equation}
The bounds on $b_{in}$ in $x y$-system dictate, $-b_c<b_{in}=-d_{BB}\le 0$. Therefore, (\ref{eqn:dBotBot}) becomes, 
$-b_c\le q\ell\cos\theta - p\ell\sin\theta < 0$. The latter can be rearranged to,
\begin{equation}
0 < \sin\left(\theta-\alpha\right) \le \frac{b_c}{s\ell},
\label{eqn:dBotBotCondn}
\end{equation}
where, $s\left([p,q]\right)=\sqrt{p^2+q^2}$.

For top-bottom-top collisions leading to periodicity, we can similarly arrive at $-b_c<p_1\ell\sin\theta - 
q_1\ell\cos\theta + b_c \le 0$ and $0\le p_2\ell\sin\theta - q_2\ell\cos\theta -b_c < b_c$ for the first 
(top-bottom) and the second (bottom-top) collisions, respectively. Since these always occur successively 
in a periodic trajectory, we can add the two inequalities to yield,
\[-b_c \le (q_1+q_2)\ell\cos\theta - (p_1+p_2)\ell\sin\theta \le b_c.\]

Using the total periodiciy $[p,q] = [p_1+p_2,q_1+q_2]$ and $s$ as defined earlier, the above inequality can 
be rearranged to take a form similar to (\ref{eqn:dTopTopCondn}) or (\ref{eqn:dBotBotCondn}): 
\[-\frac{b_c}{s\ell}\le\sin\left(\alpha-\theta\right)\le\frac{b_c}{s\ell}~~\equiv~~-\frac{b_c}{s\ell}\le\sin\left(\theta-\alpha\right)\le\frac{b_c}{s\ell}.\]

In the above double-inequalities, only one side becomes relevant depending on the relative magnitudes of 
$\theta$ and $\alpha$. If $\theta<\alpha$ we have $\sin\left(\alpha-\theta\right)>0$, and inequality 
(\ref{eqn:dTopTopCondn}) is relevant, whereas in the case of $\theta>\alpha$, the inequality (\ref{eqn:dBotBotCondn})
is the appropriate choice.

Thus, (\ref{eqn:dTopTopCondn}) and (\ref{eqn:dBotBotCondn}) together describe the periodic behavior of the 
particle trajectories in the lattice. Both can be combined into a single inequality as,
\begin{equation}
 \left|\sin\left(\alpha - \theta\right)\right| \le \frac{b_c}{s\ell}.
\label{eqn:universalIneq}
\end{equation} 
We observe that the values of $\theta$ satisfying the inequality (\ref{eqn:universalIneq}) are symmetric about 
$\theta=\alpha$. Which means, if $\tilde{\theta_c}$ and $\theta_c$ satisfy the {\em equalities} corresponding to 
(\ref{eqn:dTopTopCondn}) and (\ref{eqn:dBotBotCondn}), respectively, then \[\alpha=\frac{\tilde{\theta_c}+\theta_c}{2}.\]
Further, note that (\ref{eqn:dTopTopCondn}) and (\ref{eqn:dBotBotCondn}) are {\em necessary conditions} for periodicity 
of a trajectory in a strict mathematical sense, but they are not sufficient conditions. Which means, {\em if} 
a trajectory is known to exhibit $[p,q]$-locking, {\em then} the pair $[p,q]$ must satisfy (\ref{eqn:dTopTopCondn})
or (\ref{eqn:dBotBotCondn}) depending upon the relative magnitudes of $\alpha$ and $\theta$. Conversely, there may 
exist many integer pairs $[p,q]$ which satisfy (\ref{eqn:dTopTopCondn}) or (\ref{eqn:dBotBotCondn}), for a given 
forcing angle $\theta$ and parameters $b_c$ and $\ell$. However, physically, the trajectory would become periodic 
after a collision with the obstacle closest to the one at the origin, i.e., {\em only if} the pair $[p,q]$ 
is the closest possible pair to the origin $[0,0]$ satisfying the inequalities. Thus, the 
converse problem of finding the periodicity $[p,q]$ lies in the domain of mixed integer minimization problems, 
stated as: minimize $\sqrt{p^2+q^2}$ for integers $p$ and $q$ subject to the constraints $p>0$, $q\ge0$, and the 
inequalities (\ref{eqn:dTopTopCondn}) and (\ref{eqn:dBotBotCondn}).

\begin{figure}
\begin{center}
 \includegraphics[width=0.85\fw]{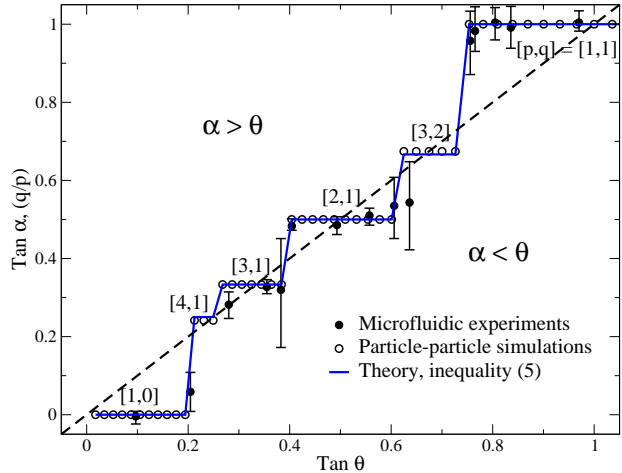}
\end{center}
 \caption{Migration direction ($\tan\alpha$) versus forcing direction ($\tan\theta$) portraying 
devil's staircase-like structure representing directional locking. The empty circles represent 
individual particle-particle simulations under the dilute approximation, the line represents the 
solution of (\ref{eqn:universalIneq}), $[p,q]$, such that the integer pair $[p,q]$ is the closest 
integer pair to $[0,0]$. The filled circles with error bars correspond to the data from microfluidic 
experiments \citep{devendra2012}. The dashed line represents the $1$-$1$-line dividing the plane in 
the regions with $\alpha>\theta$ (the region above the line) and $\alpha<\theta$ (the region below 
the line) as shown. The inequality (\ref{eqn:dTopTopCondn}) is satisfied along the solid line forming 
the staircase in the region above the $1$-$1$-line, whereas the inequality (\ref{eqn:dBotBotCondn}) 
is satisfied along the solid line forming the staircase below the $1$-$1$-line.}
 \label{fig:devilsStaircase}
\end{figure}

In figure \ref{fig:devilsStaircase}, we show an excellent agreement between the migration angles ($\tan\alpha=q/p$) 
obtained by solving either (\ref{eqn:dTopTopCondn}) or (\ref{eqn:dBotBotCondn}) for the pairs $[p,q]$ corresponding 
to the smallest $s\left([p,q]\right)$ (using {\tt Mathematica$^\text{\textregistered}$}), and those obtained from 
trajectory calculations using particle-particle simulations under the dilute assumption \citep{frechette2009}, for 
different forcing angles. The same critical offset $b_c$ is used in both cases, which corresponds to a particle of 
the same size as the obstacle ($a=b$) and a range of non-hydrodynamic interactions $\epsilon=10^{-3}a$ (the 
inter-obstacle spacing is $\ell=5a$). The same figure also shows an agreement between data from microfluidic experiments 
\citep{devendra2012} and theory (the ratio $b_c/\ell$ corresponding to the experimental data shown in the figure is 
approximately equal to that used in theoretical calculations). 

\subsection{Transitions from and to periodicities}\label{subsec:deriveTransitions}
\begin{figure*}
\includegraphics[width=1.25\fw]{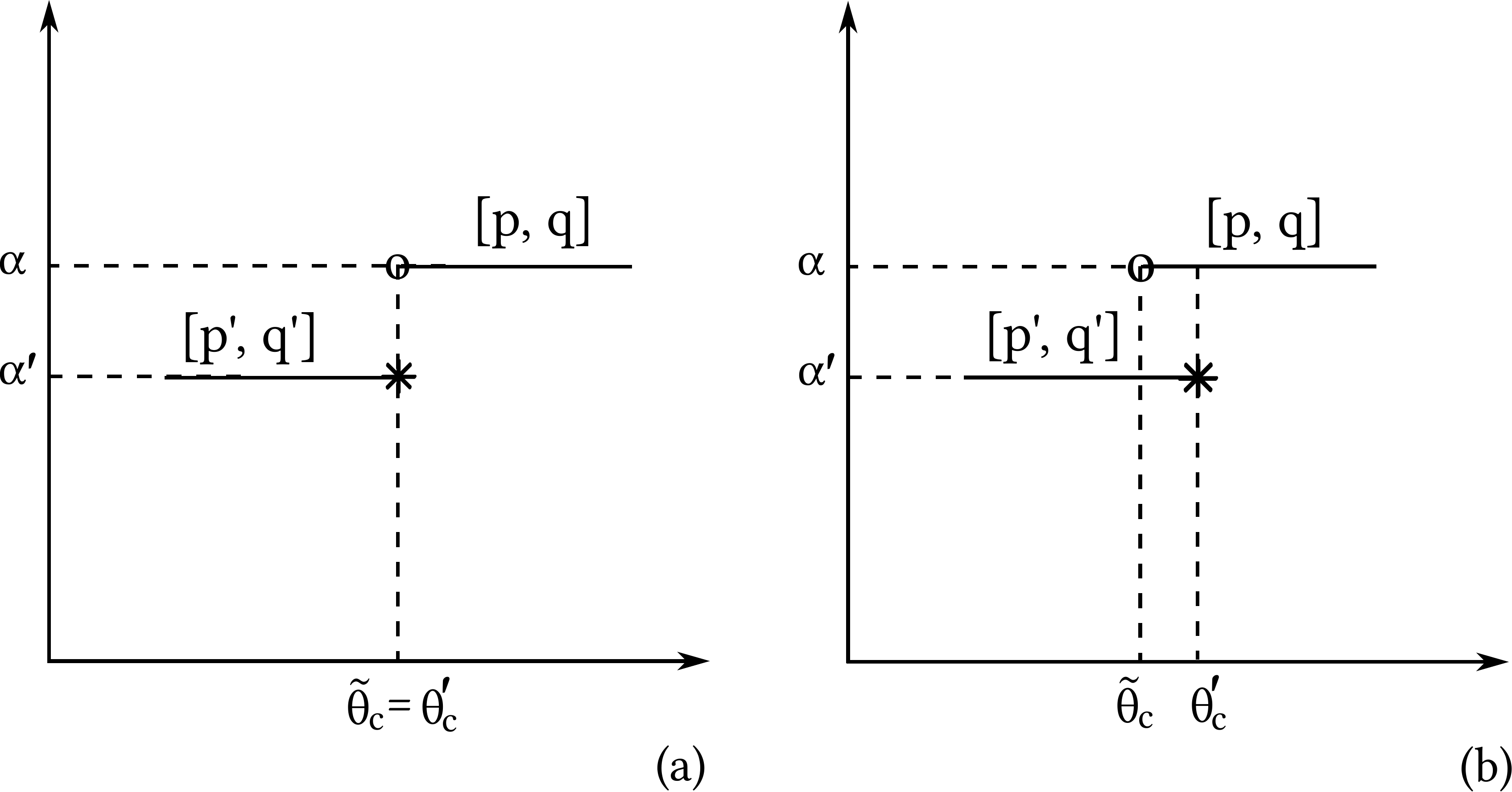}
\caption{Schematic showing two consecutive locking directions, $[p^\prime,q^\prime]$ and $[p,q]$. (a) There 
is no region in which the periodicities satisfy either of the two inequalities (\ref{eqn:dTopTopCondn}) and 
(\ref{eqn:dBotBotCondn}) simultaneously. (b) Both periodicities satisfy one of the two inequality simultaneously 
in the region of overlap. The transition angles at the end of $[p^\prime,q^\prime]$ and beginning of $[p,q]$ 
are denoted by $\theta_c^\prime$ and $\tilde{\theta_c}$, respectively.}
\label{fig:transitionSchematic}
\end{figure*}

The transition-forcing-angles from one locked migration direction to the next as well as the migration 
angle itself, can be computed by treating (\ref{eqn:dTopTopCondn}) and (\ref{eqn:dBotBotCondn}) 
as equalities. We have noted earlier that for each periodicity $[p,q]$ corresponding to a migration 
angle $\tan\alpha=q/p$, two distinct transition angles $\tilde{\theta_c}$ and $\theta_c$ 
can be obtained from the equalities corresponding to (\ref{eqn:dTopTopCondn}) and (\ref{eqn:dBotBotCondn}), 
by solving $\sin(\alpha-\tilde{\theta_c})=b_c/s\ell$ and $\sin\left(\theta_c-\alpha\right)=b_c/s\ell$, 
respectively. Then, consider two consecutive locking directions $[p^\prime,q^\prime]$ and $[p,q]$, with 
the primed direction representing the `lower' step in the staircase (i.e., $\alpha' < \alpha$), as shown 
in schematic \ref{fig:transitionSchematic}. 
If the periodicities do not {\em simultaneously} satisfy the inequalities (\ref{eqn:dTopTopCondn}) and 
(\ref{eqn:dBotBotCondn}), then there is no overlap between the two steps as shown in figure 
\ref{fig:transitionSchematic}(a). Therefore, the the transition from $[p^\prime,q^\prime]$ to $[p,q]$ 
takes place at the end of the  lower step, given by $\theta=\theta_c^\prime$ (`*' point in the figure). 
But since there is no overlap between the steps, this has to be the angle at the beginning of the $[p,q]$-step, 
given by $\theta=\tilde{\theta_c}$ (`o' point in the figure). Therefore, this is the case when both critical angles 
are equal ($\tilde{\theta_c} = \theta_c^\prime$), and either equality corresponding to 
(\ref{eqn:dTopTopCondn}) or (\ref{eqn:dBotBotCondn}) gives the same result. An example of such a 
transition is seen in figure \ref{fig:devilsStaircase} from $[3,1]$-periodicity to $[2,1]$-periodicity. 
Although $s([2,1])=\sqrt{5} < s([3,1]) = \sqrt{10}$, there is no forcing direction $\theta$ for which both 
inequalities are satisfied, and there is no overlap between the corresponding steps.
If there is an overlap between the steps as shown in figure \ref{fig:transitionSchematic}(b), the two 
inequalities are satisfied for the forcing directions $\theta$ in the overlap region. Then in the region of overlap, 
the step with a smaller $s$ is realised, thus satsfying the physical requirement that the trajectory becomes 
periodic with the shortest period. An example of this type of transition is that from $[4,1]$-periodicity to 
$[3,1]$-periodicity, shown in figure \ref{fig:devilsStaircase}. In this case, since $s([3,1]) = \sqrt{10} < 
s([4,1]) = \sqrt{17}$, the transition occurs before $\theta=\theta_{c,[4,1]}$ (the end of $[4,1]$-step). 
Therefore, in the case of an overlap between the steps, 
\begin{itemize}
 \item[(a)] if $s < s^\prime$, then the transition occurs at the beginning of $[p,q]$ (equality in (\ref{eqn:dTopTopCondn})) at the `o'-point
(in figure \ref{fig:transitionSchematic}(b))
 \item[(b)] else/otherwise, the transition occurs at the end of $[p^\prime,q^\prime]$ at the `*'-point in the figure.
\end{itemize}
We apply the above argument to the transitions from- and to- $[1,0]$ and $[1,1]$-directions, respectively the 
simplest possible locking directions.
\begin{itemize}
 \item The transition from $[1,0]$: Since $[1,0]$ gives the smallest possible $s$-value ($s=1$), the transition 
always occurs at the end of $[1,0]$-step, i.e., using the equality in (\ref{eqn:dBotBotCondn}),\label{subsubsec:firstTranAngle}
\begin{equation}
 \sin\left(\theta_F-0\right)=\sin\theta_F=\frac{b_c}{\ell},
\label{eqn:firstTranAngle}
\end{equation}
where $\theta_F$ is defined as the first transition angle. 
 \item The transition to $[1,1]$: The final locking direction $[1,1]$ gives the second smallest possible 
$s$-value ($s=\sqrt{2}$). Therefore, the transition to $[1,1]$-direction always occurs at the beginning 
of the $[1,1]$-step, i.e., using the equality in (\ref{eqn:dTopTopCondn}),\label{subsubsec:lastTranAngle}
\begin{equation}
 \sin\left(\frac{\pi}{4} - \theta_L\right) = \frac{b_c}{\sqrt{2}\ell},
\label{eqn:lastTranAngle}
\end{equation}
where, $\theta_L$ is the last transition angle.
\end{itemize}
The only exception to (\ref{eqn:lastTranAngle}) is when there is a direct transition from $[1,0]$ to $[1,1]$, 
which is the case of a one-step staircase. In this case, 
$s = 1 < s = 2$, the transition takes place at $\theta=\theta_F$ (the end of the $[1,0]$-step), which may 
or may not be the same as $\theta=\theta_L$ (the beginning of the $[1,1]$-step). 
\subsubsection{The first locking direction after $[1,0]$}\label{subsubsec:firstLockDirP}
If the locking direction after the first transition is $[p_F,q_F]$, then 
\[\left|q_F\cos\theta_F-p_F\sin\theta_F\right|\le\frac{b_c}{\ell}.\] 
However, from (\ref{eqn:firstTranAngle}), $\sin\theta_F=b_c/\ell$. Also, increasing $\theta$ counter-clockwise 
from $X$- to $Y$-axis, the first transition should be from a locking direction along the zeroth row of 
obstacles along $X$-axis (i.e., $[1,0]$) to the first row of obstacles along $X$-axis (i.e., $[p,1]$ for some 
integer $p$). Therefore, $q_F=1$ (see figure \ref{fig:systemGeometryDLD}(b)). 
Thus, $\left|\cot\theta_F-p_F\right|\le 1$. Since $p_F$ is an integer, we get, 
\begin{equation}
 p_F=\left\lfloor\cot\theta_F\right\rfloor~~\ldots~~\left\lfloor.\right\rfloor\equiv~\text{floor function}
\label{eqn:firstLockDirP}
\end{equation} 
Thus, the locked direction after the first transition is $\tan\alpha=q_F/p_F=1/\left\lfloor\cot\theta_F\right\rfloor$, 
where $\theta_F$ is given by (\ref{eqn:firstTranAngle}) above.
\subsubsection{The last locking direction before $[1,1]$}\label{subsubsec:lastLockDirQ}
If the locking direction before the final transition is $[p_L,q_L]$, then 
\[\left|q_L\cos\theta-p_L\sin\theta\right|\le \frac{b_c}{\ell}.\] 
From (\ref{eqn:lastTranAngle}), $\cos\theta_L-\sin\theta_L = b_c/\ell$. Further, increasing $\theta$ counter-clockwise 
from $X$- to $Y$-axis, the last transition from $[p_L,q_L]$ to $[q_L,q_L]$ (i.e., $[1,1]$) should satisfy 
$p_L = q_L + 1$. Thus, the above inequality becomes, 
$\left|\left(\frac{\sin\theta_L}{\cos\theta_L-\sin\theta_L}\right)-q_L\right|\le 1$. Since $q_L$ is an integer,
\begin{equation}
 q_L = \left\lfloor\frac{\sin\theta_L}{\cos\theta_L-\sin\theta_L}\right\rfloor = \left\lfloor\frac{\tan\theta_L}{1-\tan
\theta_L}\right\rfloor.
\label{eqn:lastLockDirQ}
\end{equation}
Therefore, the locked direction before the final transition to $[1,1]$ is given as 
\[\tan\alpha = q_L/p_L = \frac{\left\lfloor\frac{\tan\theta_L}{1-\tan\theta_L}\right\rfloor}{\left\lfloor\frac{
\tan\theta_L}{1-\tan\theta_L}\right\rfloor+1},\] where, $\theta_L$ is the solution of (\ref{eqn:lastTranAngle}).
We note again, that the only exception to the calculation leading to (\ref{eqn:lastLockDirQ}) is the case when 
the transition to $[1,1]$ occurs from $[1,0]$. In this case, the transition occurs at corresponding $\theta_F$ 
instead of $\theta_L$.

\section{Design rules and separation resolution in DLD for simple staircase structures}\label{sec:designRules}
\begin{figure}
\begin{center}
 \includegraphics[width=0.85\fw]{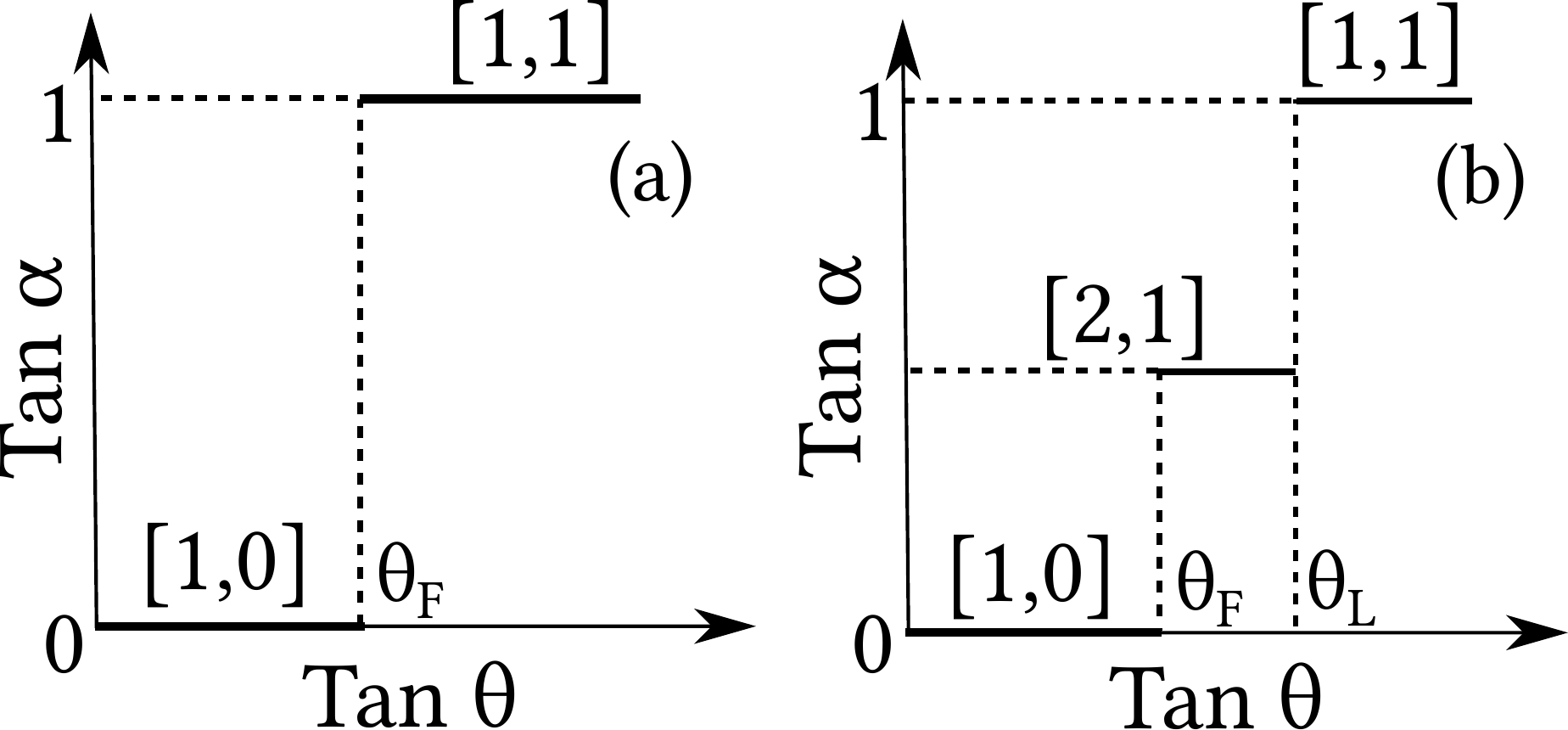}
\end{center}
 \caption{(a) 1-step staircase with transition $[1,0]\mapsto[1,1]$ at $\theta_F$ (b) 2-step staircase $[1,0]\mapsto[2,1]
\mapsto[1,1]$ with transitions at $\theta_F$ and $\theta_L$.}
 \label{fig:twoStaircases}
\end{figure}

We first derive the constraints on the ratio $b_c/\ell$ for a particle to exhibit exactly one transition (figure 
\ref{fig:twoStaircases}(a)) and exactly two transitions (figure \ref{fig:twoStaircases}(b)), based on 
(\ref{eqn:universalIneq}) and the discussion in \S\ref{sec:dirLock}. 
We have shown in \S\ref{subsec:deriveTransitions} that the first transition angle is given by $\sin\theta_F=b_c/\ell$ 
corresponding to the transition from the $[1,0]$ locking direction. Further, we have noted earlier that $\theta_L$ from 
(\ref{eqn:lastLockDirQ}) is not necessarily equal to $\theta_F$, since the transition to $[1,1]$ from $[1,0]$ is an 
exception. Therefore, applying the constraint (\ref{eqn:dTopTopCondn}) for $\theta_F$ we get,
\begin{equation}
 \sin\left(\frac{\pi}{4}-\theta_F\right)\le\frac{b_c}{\ell\sqrt{2}}\Rightarrow\tan\theta_F\le\frac{1}{2},
\label{eqn:oneStepTheta}
\end{equation}
and,
\begin{equation}
 \frac{b_c}{\ell} \ge \frac{1}{\sqrt{5}}~\ldots~\left(\text{using }\sin\theta_F=\frac{b_c}{\ell}\right).
\label{eqn:oneStepbc}
\end{equation}
Thus, for a fixed particle radius $a$, obstacles of size $b$, and a range $\epsilon$ of non-hydrodynamic interactions 
(i.e., for a fixed $b_c$), a square lattice with $\ell \ge \sqrt{5} b_c$ can be constructed in which the particle
exhibits a 1-step staircase structure.

In the case of a 2-step staircase (figure \ref{fig:twoStaircases}(b)), the locking direction after the first 
transition from $[1,0]$ is $[2,1]$, which is the same as the locking direction before the final transition to 
$[1,1]$. Using \S\S\ref{subsubsec:firstLockDirP}, \ref{subsubsec:lastLockDirQ},
\[\tan\alpha=\frac{1}{2}=\frac{1}{\left\lfloor\cot\theta_F\right\rfloor}=\frac{\left\lfloor\frac{\tan\theta_L}{1-
\tan\theta_L}\right\rfloor}{\left\lfloor\frac{\tan\theta_L}{1-\tan\theta_L}\right\rfloor+1}~~\&~~~\sin\theta_F=b_c/\ell\]

The first set of equations above yields, $\left\lfloor\cot\theta_F\right\rfloor=2$ and $\left\lfloor\frac{\tan\theta_L}{
1-\tan\theta_L}\right\rfloor=1$. Thus, using $\sin\theta_F=b_c/\ell$, we obtain, 
\begin{align}
 \frac{1}{3}<\tan\theta_F&\le\frac{1}{2}\le\tan\theta_L<\frac{2}{3} \label{eqn:twoStepTheta}\\
 \frac{1}{\sqrt{10}}&<\frac{b_c}{\ell}\le\frac{1}{\sqrt{5}} \label{eqn:twoStepbc}.
\end{align}

Thus, if a square lattice satisfies (\ref{eqn:twoStepbc}) for a particle of radius $a$, a critical parameter $b_c$ 
(a function of $a$, $b$ and $\epsilon$) and unit cell $\ell$, then the particle exhibits locking with a 2-step 
staircase structure with the corresponding two transition angles satisfying (\ref{eqn:twoStepTheta}).

\subsection{Design constraints and separation resolution in DLD}\label{subsec:desConOpt}
In pairwise size-based separation, two particles of different sizes, say $a$ and $a^\prime$ (and perhaps 
different length-scales corresponding to the range of non-hydrodynamic interactions, say $\epsilon$ 
and $\epsilon^\prime$) exhibit two distinct critical parameters, viz.- $b_c$ and $b_c^\prime$. 
Separation is possible at forcing angles such that the migration directions $\tan\alpha=q/p$ and 
$\tan\alpha^\prime=q^\prime/p^\prime$ are distinct. Further, a larger difference between the 
migration directions is synonymous with a higher resolution. Thus, a simple design strategy is 
to maximize $\left|\alpha-\alpha^\prime\right|$. 

By rearranging (\ref{eqn:universalIneq}) as,
\begin{equation}
 \left|\sqrt{q^2+p^2}\sin\left(\alpha-\theta\right)\right|\approx\left|\sqrt{q^2+p^2}\left(
\alpha-\theta\right)\right|\le\frac{b_c}{\ell},
\label{eqn:genericIneq}
\end{equation}
where, the last inequality results from the small angle approximation $\sin(\alpha-\theta)\approx(
\alpha-\theta)$ since $0<\alpha,\theta\le\pi/4$. Similarly, a good approximation for 
$\left|\alpha-\alpha^\prime\right|$ can be obtained by combining (\ref{eqn:genericIneq}) 
for both particles $a\text{ and }a^\prime$,
\begin{equation}
\left|\alpha-\alpha^\prime\right|\le\frac{b_c}{\left(\sqrt{p^2+q^2}\right)\ell}+\frac{b_c^\prime}{\left(
\sqrt{p^{\prime 2}+q^{\prime 2}}\right)\ell}
\label{eqn:approxDeviationFromForcing}
\end{equation}

As an immediate consequence of (\ref{eqn:genericIneq}), 
we note that the largest difference between the migration direction and the forcing direction (i.e., 
$\left|\tan\alpha-\tan\theta\right|\approx\left|\alpha-\theta\right|$) occurs before the first transition 
from $\alpha=0$ ($\text{locking}\equiv[1,0]$ and when $p^2+q^2=1$), and it is always equal to 
$\left|\alpha-\theta\right|=\theta\lesssim b_c/\ell$. Furthermore, along with (\ref{eqn:approxDeviationFromForcing}), 
we infer that the largest separation resolution between two species can be obtained when one of the 
species has undergone its first transition, while the other is still locked in $[1,0]$ direction (for 
example, $\theta_F<\theta<\theta_F^\prime$). This observation supports our earlier experimental 
inference that, it is the most beneficial strategy to set the forcing angle between the first transitions 
($\theta_F$ and $\theta_F^\prime$) of the two species undergoing separation 
\citep{balvin2009,bowman2012,devendra2012}.
The inequality, i.e., expression (\ref{eqn:approxDeviationFromForcing}) not only gives an upper bound on the resolution, but also 
gives design constraints on the obstacle radius $b$ and the lattice spacing $\ell$ through the ratio 
$b_c/\ell$ for known locking directions ($[p,q]$ and $[p^\prime,q^\prime]$), a fixed forcing angle $\theta$ 
and known radii of particles ($a$ and $a^\prime$). In the following, we illustrate this result for particles 
with simple staircase structures, viz.- only one transition from $[1,0]$ to $[1,1]$ and two transitions 
$[1,0]\mapsto[2,1]\mapsto[1,1]$.

For a mixture of two species, both exhibiting 1-step staircases with different transitions $\theta_F$ and 
$\theta_F^\prime$ satisfying (\ref{eqn:oneStepTheta}) and (\ref{eqn:oneStepbc}), it is readily understood 
that the forcing angle needs to be between these two values 
if any separation is desired. The separation resolution $\left|\alpha-\alpha^\prime\right|$ is always $\pi/4$ 
in this case, since one species is always locked in $[1,0]$-periodicity, while the other is locked in 
$[1,1]$-periodicity for a forcing angle between $\theta_F$ and $\theta_F^\prime$.

\begin{figure*}
\begin{center}
 \includegraphics[width=1.35\fw]{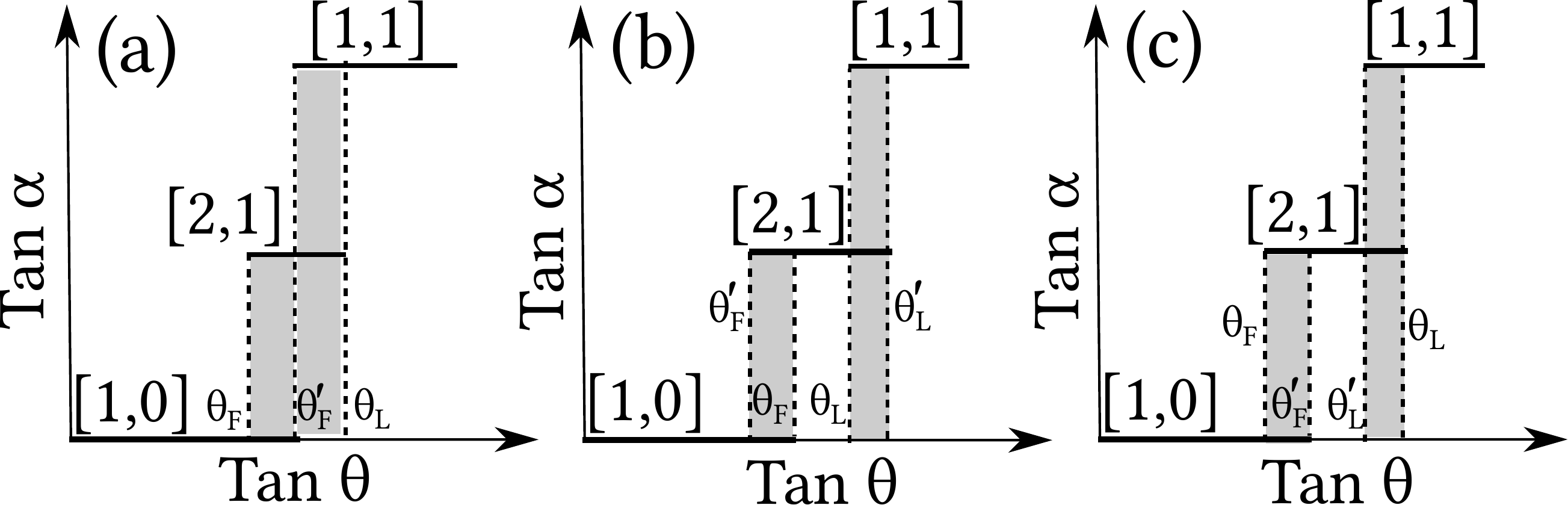}
 \caption{Three combinations of the simplest staircase structures possible: (a) one particle exhibits 1-step staircase, 
the other exhibits 2-step staircase, with the former transition lying between the two transitions of the latter, (b) and (c) 
both particles exhibit 2-step staircase structures. The shaded areas represent the ranges of forcing angle $\theta$ in which 
separation between the two species is possible.}
 \label{fig:separationOptimization}
\end{center}
\end{figure*}

In the case of a mixture of a species exhibiting a 1-step staircase (say, for particles of radius $a^\prime$) and 
another species exhibiting a 2-step staircase (say, for particles of radius $a$), then (\ref{eqn:oneStepTheta}) and 
(\ref{eqn:twoStepTheta}) permit only one possible scenario depicted in (figure \ref{fig:separationOptimization}(a)). 
After appropriate algebra corresponding to this case, we get,
\[\frac{1}{3}<\tan\theta_F\le\frac{1}{2}\le\tan\theta_F^\prime\le\tan\theta_L<\frac{2}{3}.\]
Similarly, for a mixture of particles exhibiting a 2-step staircase, (\ref{eqn:firstTranAngle}), 
(\ref{eqn:lastTranAngle}) and (\ref{eqn:twoStepTheta}) permit only two cases depicted in figures 
\ref{fig:separationOptimization}(b) and \ref{fig:separationOptimization}(c), which correspond to the two 
cases $b_c>b_c^\prime$ or $b_c<b_c^\prime$, respectively.
Again, applying these constraints we get,
\[\frac{1}{3}<\tan\theta_F^\prime<\tan\theta_F\le\frac{1}{2}<\tan\theta_L<\tan\theta_L^\prime<\frac{2}{3},\] 
or \[\frac{1}{3}<\tan\theta_F<\tan\theta_F^\prime\le\frac{1}{2}<\tan\theta_L^\prime<\tan\theta_L<\frac{2}{3}.\]

In terms of separation, it is evident from the figure that separation 
between primed and non-primed species is possible only if the forcing angle satisfies 
\begin{itemize}
 \item[(i)] $\theta\in[\theta_F,\theta_F^\prime]\cup[\theta_F^\prime,\theta_L]$ corresponding to 
figure \ref{fig:separationOptimization}(a), 
 \item[(ii)] $\theta\in[\theta_F^\prime,\theta_F]\cup[\theta_L,\theta_L^\prime]$ for figure \ref{fig:separationOptimization}(b),
 and 
 \item[(iii)]$ \theta\in[\theta_F,\theta_F^\prime]\cup[\theta_L^\prime,\theta_L]$ in the case of figure \ref{fig:separationOptimization}(c).
\end{itemize}
Further, the figure also indicates that $\left(\pi/4-\arctan\left(1/2\right)\right)$ and $\arctan\left(1/2\right)$ 
are the only two separation resolutions $\left(\left|\alpha-\alpha^\prime\right|\right)$ corresponding to these cases.

Thus, the maximum separation resolution between species corresponding to the three cases shown in figure \ref{fig:separationOptimization} 
is $\arctan(1/2)\approx26.56^\circ$ since it is greater in magnitude than $[\pi/4 - \arctan(1/2)]$, and it occurs 
if $\theta\in[\theta_F,\theta_F^\prime]$ for figure \ref{fig:separationOptimization}(a), 
$\theta\in[\theta_F^\prime,\theta_F]$ for figure \ref{fig:separationOptimization}(b) and 
$\theta\in[\theta_F,\theta_F^\prime]$ for figure \ref{fig:separationOptimization}(c). As highlighted in the context 
of (\ref{eqn:approxDeviationFromForcing}), this conclusion is consistent with our experimental observation, that the 
forcing angle between the first transition angles of the species to be separated, achieves the best resolution 
\citep{balvin2009,bowman2012,devendra2012}.

\section{Summary}\label{sec:summaryDLD}
In summary, we have presented a theoretical analysis of the directional locking phenomenon exhibited 
by particles navigating through a square array of obstacles, in the limit of negligible particle and 
fluid inertia. In the dilute limit for the array (i.e., a sparse array), interactions between a single 
obstacle and a particle are sufficient for trajectory analysis. Coupled with the dilute assumption, we 
have used a critical parameter (incorporating both hydrodynamic as well as short-range repulsive 
non-hydrodynamic particle-obstacle interactions) to replace the physical particle-obstacle system with 
its kinematically equivalent abstraction. Within the abstract model, the particle is replaced by a 
point-particle, while the obstacle radius is scaled to be equal to the critical parameter. Due to the 
model, a simple geometric analysis suffices to derive the periodicity condition, both necessary and sufficient 
for the particle trajectory. The periodicity condition directly leads to the devil's-staircase-like 
behavior of the migration direction as a function of the forcing direction. Further, using the 
periodicity condition, we have computed the design constraints on the ratio of the critical parameter to 
the lattice spacing of the square array, and commented on the resolution of deterministic separations, 
when the particle exhibits simple staircase structures. 

This work was partially supported by the National Science Foundation 
Grant Nos. CBET-0731032 and CBET-1339087.

\end{document}